\documentclass[aps,prl,twocolumn,showpacs,floatfix]{revtex4}

\usepackage{epsfig}

\usepackage[dvips]{color}
\definecolor{Black}{named}{Black}
\definecolor{Red}{named}{Red}

\def\d{{\rm d}}
\def\be{\begin{equation}}
\def\ee{\end{equation}}
\def\bu{{\mathbf u}}
\def\bp{{\mathbf p}}
\def\bpp{{\mathbf p}^\prime}
\def\pp{p^\prime}
\def\fp{f^\prime}
\def\Ep{E^\prime}
\def\Ip{I^\prime}
\def\d{{\rm d}}
\def\be{\begin{equation}}
\def\ee{\end{equation}}
\def\lsim{\raise0.3ex\hbox{$\;<$\kern-0.75em\raise-1.1ex\hbox{$\sim\;$}}}
\def\gsim{\raise0.3ex\hbox{$\;>$\kern-0.75em\raise-1.1ex\hbox{$\sim\;$}}}
\def\dCG{{A}_{\rm CCG}}
\def\bu{{\mathbf u}}
\def\bp{{\mathbf p}}
\def\br{{\mathbf r}}
\def\bpp{{\mathbf p}^\prime}
\def\brp{{\mathbf r}^\prime}
\def\pp{p^\prime}
\def\fp{f^\prime}
\def\Ep{E^\prime}
\def\Ip{I^\prime}

\begin{document}

\title{The Compton-Getting effect on ultra-high energy cosmic rays of
  cosmological origin}

\author{M.~Kachelrie{\ss}$^1$, P.~D.~Serpico$^2$}
\affiliation{$^{1}$Institutt for fysikk, NTNU Trondheim, N--7491
  Trondheim, Norway
  \\$^{2}$Max-Planck-Institut f\"ur
  Physik (Werner-Heisenberg-Institut), D--80805 Munich, Germany}

\date{\today}

\begin{abstract}
Deviations from isotropy have been a key tool to identify the origin
and the primary type of cosmic rays at low energies. We suggest that
the Compton-Getting effect can play a similar role at ultra-high
energies: If at these energies the cosmic ray flux is dominated by
sources at cosmological distances, then the movement of the Sun
relative to the cosmic microwave background frame induces a dipole
anisotropy at the 0.6\% level. The energy dependence and the
orientation of this anisotropy provide important information about
the transition between galactic and extragalactic cosmic rays, the
charge of the cosmic ray primaries, the galactic magnetic field and,
at the highest energies, the energy-loss horizon of cosmic rays. A
$3\sigma$ detection of this effect requires around $10^6$ events in
the considered energy range and is thus challenging but not
impossible with present detectors. As a corollary we note that the
Compton-Getting effect allows one also to constrain the fraction of
the diffuse $\gamma$-ray background emitted by sources at
cosmological distance, with promising detection possibilities for
the GLAST satellite.
\end{abstract}
\pacs{96.50.S-,
98.70.Vc
\hfill MPP-2006-50}

\maketitle

\section{Introduction}%
Ultra-high energy cosmic ray (UHECR) physics has gained increasing
momentum in recent years.  While the present state of observations
is still puzzling~\cite{reviews}, new experiments like the Pierre
Auger Observatory (PAO)~\cite{PAO} or the Telescope Array
(TA)~\cite{TA} are expected to shed light on many unresolved issues
with their improved detection techniques and increased statistics.

Among the most important open questions are the origin and the
composition of UHECRs. Cosmic rays with energy below $E\sim
10^{16}$~eV are generally believed to be accelerated in galactic
supernova remnants, but at higher energies their sources are
unknown. Given the strength of the galactic magnetic field (GMF),
one would expect a significant excess of events towards the galactic
plane at least at energies above 10$^{19}$~eV if these CRs have a
galactic origin. Since up to the highest energies the arrival
directions of CRs show no correlation with the galactic plane, the
UHE part of the spectrum is generally thought to be extragalactic.
Moreover, Hillas' argument~\cite{Hillas} that the Larmor radius of
an accelerated particle should fit inside the accelerator favors
extragalactic sources as origin of the cosmic rays with the highest
energies.

At present, two main models exist for the transition between
galactic and extragalactic sources: The first one argues that the
ankle in the cosmic ray spectrum at $5\times10^{18}$~eV is caused by
the cross-over from the steep end of the galactic to the flatter
extragalactic flux~\cite{ankle}. The second one interprets the ankle
as dip produced by $e^+e^-$ pair production of extragalactic protons
with CMB photons~\cite{dip}. Then the transition between between
galactic and extragalactic CRs could take place at energies as low
as a ${\rm few}\times 10^{17}$~eV~\cite{Berezinsky:2004wx}. An
extreme but not firmly excluded possibility is that all cosmic rays
are galactic as, e.g., in the Zevatron model~\cite{ZeV}. In this
case, the UHECR primaries should be heavy nuclei because the GMF can
isotropize them only sufficiently---if at all---for a large electric
charge $Qe$.

Extensive air shower experiments can in principle measure the
chemical composition of the UHECR flux and thus determine the
transition from a galactic, iron-dominated  component to
extragalactic protons. However, the uncertainties in the hadronic
interaction models become so large above $E\sim 10^{17}\,$eV that a
reliable differentiation between proton and nuclei primaries is at
present impossible~\cite{chemie}. Moreover this method fails if the
extragalactic component is also dominated by heavy nuclei.

Anisotropies in the CR flux are another important tool to
distinguish between different origin and primary models. Theoretical
predictions of anisotropies for galactic sources depends on the GMF
and the exact source distribution. The amplitude $A$ of galactic
anisotropies increases with energy and may range from $A\sim
10^{-4}$ at $E\sim\,{\rm few}\,\times 10^{14}$~eV to $A\sim 10^{-2}$
at $E\sim\,{\rm few}\,\times 10^{17}$~eV~\cite{Candia03}. By
contrast, in most analyses the extragalactic flux is assumed to be
isotropic. However, already Ref.~\cite[p.~160]{Ber90} noticed that
the movement of the Sun relative to the cosmic microwave background
frame induces a dipole anisotropy at the 0.6\% level. The
experimental data at that time indicated a larger anisotropy at
$E\sim 10^{17}$~eV. Furthermore, it was believed that extragalactic
protons dominate the CR flux only at $E\gsim 10^{19}$~eV. As a
result, this idea was not followed up.

A similar effect connected with the rotation of the Milky Way was
proposed already 70 years ago by Compton and Getting, and was
recognized as a diagnostic tool for low-energy cosmic
rays~\cite{CGpaper}. More recently, the importance of the
cosmological Compton-Getting (CCG) effect has been stressed as a
signature for the cosmological origin of gamma ray
bursts~\cite{Maoz:1993uq}. An analysis of its potential as a
diagnostic tool in UHECR physics is however, to the best of our
knowledge, still missing. In the following, we shall perform such an
analysis. We shall argue that the CCG provides information about the
transition between galactic and extragalactic cosmic rays, their
charge, the GMF, and, at the highest  energies, the energy-loss
horizon of cosmic rays. We also briefly discuss the CCG effect on
the diffuse  $\gamma-$ray background, and comment on the chances for
a detection of the two signatures in forthcoming experiments.

\section{The cosmological Compton-Getting effect}%
Let us recall briefly the derivation of the Compton-Getting effect
(see e.g.~\cite[p.~30]{Ber90}). An observer in motion with velocity
$\bu$ relative to the coordinate system in which the distribution of
cosmic rays is isotropic will measure an anisotropic cosmic ray
flux. If UHECR sources are on average at rest with respect to the
cosmological frame, the magnitude and direction of the velocity
$\bu$ of the solar system can be deduced from the detection of the
dipole anisotropy in the CMB,  $u=368\pm 2$~km/s in the direction
$(l,b)=(263.85^\circ,48.25^\circ)$~\cite{Eidelman:2004wy}. Since
$u\equiv|\bu|\ll 1$, the anisotropy induced by the CCG effect is
dominated by the lowest moment, i.e. its dipole moment.

To derive the amplitude of this anisotropy, we compare the phase
space distribution function $f$ in the frame of the observer,
denoted by $f'(\brp,\bpp)$, with the one in the frame in which the
UHECR flux is isotropic, $f(\br,\bp)$. Lorentz invariance requires
$f(\br,\bp)=\fp(\brp,\bpp)$. Using $\bp-\bpp\simeq p\,\bu$ valid for
ultra-relativistic particles and for $u\ll 1$ and suppressing from
now on the variable $\br$, we expand the phase space distribution
function in the frame of the observer,
\begin{equation}
\fp(\bpp) \simeq f(\bpp)+p\,\bu\cdot\frac{\partial
f(\bpp)}{\partial\bpp} = f(\bpp)\left(1+\frac{\bu \cdot
\bp}{p}\frac{\d \ln f}{\d\ln\pp}\right) .
\end{equation}

Cosmic ray experiments present their measured energy spectrum
normally as differential intensity $I(E)$, i.e. the number of
particles per unit solid angle and unit energy that pass per unit of
time through an area perpendicular to the direction of observation.
With $I(E)\simeq I(p)= p^2f(p)$, one obtains
\begin{equation}
\Ip(\Ep)\simeq I(E) \left[1-\left(2-\frac{\d\ln I}{\d\ln
\Ep}\right)\frac{\bu \cdot \bp}{p}\right].
\end{equation}
Thus the dipole anisotropy due to the CCG effect has the amplitude
\begin{equation}
\dCG\equiv \frac{I_{\rm max}-I_{\rm min}}{I_{\rm max}+I_{\rm
min}}=\left(2-\frac{\d\ln I}{\d\ln E}\right)\,u
\,.\label{Ianisamplitude}
\end{equation}
Taking into account the observed spectrum $I(E)\propto E^{-2.7}$ of
cosmic rays above the ankle, $\dCG=(2+2.7)\,u\simeq 0.6\%$. Note
that the Earth motion with respect to the solar system barycenter
only induces a subleading (8\%) modulation in the vector $\bu$.

\section{CCG effect and UHECRs}%
The flux of extragalactic UHECRs is isotropic in the cosmic
microwave background frame at energies $E\lsim E_\ast$ for which the
energy-loss horizon $\lambda_{\rm hor}$ of CRs is large compared to
the scale of inhomogeneities in their source distribution. In the
same energy range, peculiar velocities average out on cosmological
scales and the UHECR flux is thus isotropic at leading order. The
exact value of $E_\ast$ depends both on the density of the CR
sources and on the primary type, but $E_\ast\lsim 4\times
10^{19}\,$eV is a conservative estimate. Indeed, for protons
$\lambda_{\rm hor}$ is at the Gpc scale at $E\lsim 10^{19}\,$eV,
decreasing to about 600~Mpc at $4\times 10^{19}$~eV due to the onset
of the pion production on the CMB, and rapidly dropping to few tens
of Mpc at larger energies. For iron nuclei, $\lambda_{\rm hor}$
abruptly drops below the Gpc scale only at $E\sim 10^{20}\,$eV when
photo-dissociation processes on the microwave and infrared
backgrounds are kinematically allowed. For typical UHECRs source
densities of $n_s={\rm few}\times 10^{-5}$~Mpc$^{-3}$~\cite{ns}, the
number $N_s$ of sources contributing to the observed flux can be
estimated as (we neglect cosmological effects)
\begin{equation}
N_s\simeq \frac{4\pi}{3}\lambda_{\rm hor}^3 n_s\simeq 4.2\times 10^4
\:\frac{n_s}{10^{-5}\,{\rm Mpc}^{-3}}\, \left(\frac{\lambda_{\rm
hor}}{\rm Gpc}\right)^3 \,.
\end{equation}
Since Poisson fluctuations in $N_s$ are roughly at the 0.5\% level,
one might wonder if the CCG effect could be mimicked by a
fluctuation in the number of source per hemisphere. However, as long
as extragalactic magnetic fields wash-out anisotropies, the dominant
intrinsic fluctuation is due to the number of events $N$ observed at
the Earth and not to $N_s$, even for relatively low $N_s$. A naive
calculation of the root-mean square deflection $\theta_{\rm rms}$ of
a particle traveling the distance $L$ in a random field with
coherence scale $L_c$ and strength $B_{\rm rms}$ gives
\begin{equation}
 \theta_{\rm rms} = \frac{QeB_{\rm rms}}{E}\sqrt{\frac{L L_c}{2}}
 \simeq 120^{\circ}\,Q\,\frac{10^{19}\,{\rm eV}}{E}
        \frac{B_{\rm rms}}{{\rm nG}}
        \sqrt{ \frac{L}{\rm Gpc} \, \frac{L_c}{{\rm Mpc}} }
        \,.\label{thetarms}
\end{equation}
In the simulation performed in~\cite{Sigl:2004yk}, a significant
fraction of all UHECRs suffer deflections comparable or larger than
given by Eq.~(\ref{thetarms}), while the simulation performed
in~\cite{Dolag:2004kp} favor considerably smaller values. If the
results of the latter simulation are closer to reality, deflections
in extragalactic magnetic fields may be negligible at least for
protons even at relatively low energies such as $4\times
10^{19}$~eV. In any case, one might test observationally that
$E_\ast$ was chosen low enough by: $i)$ the approximate alignment of
the dipole axis of $\dCG$ with the one in the CMB; $ii)$ the absence
of higher multipole moments in the observed maps: While fluctuations
in the number of cosmological sources should lead to higher
multipole modes $l>1$ with similar intensity $A^{(l)}$, they are
suppressed by powers of $u$ in the case of the CCG effect,
$\dCG^{(l)}\propto u^l$.

In the following, we shall consider the statistics collected by an
experiment as the main limiting factor for the detection of the CCG
effect. It is clear that the detection of an anisotropy at the 1\%
level requires also a thorough control of systematic errors and is
therefore challenging for UHECR experiments also in this respect.

\section{Signatures and diagnostic potential for UHECRs}%
In the following, we discuss the signatures and the potential of the
CCG effect in resolving some long-standing puzzles in UHECR physics.

{\it (i)} The amplitude $\dCG$ of the anisotropy is charge- and
energy-independent, as long as the UHECR flux in the energy range
studied is dominated by sources at cosmological distance.

{\it (ii)} Since the CCG effect is a dipole anisotropy, the
magnitude of its amplitude should be robust against deflections of
UHECRs in the GMF, and only the dipole axis is displaced. The
expected deviation of the dipole vector of the CCG anisotropy from
the one in the CMB is around $20^\circ \times 10^{19}{\rm
  eV}(Q/E)$~\cite{Kachelriess:2005qm}.
Thus at energies 2--3$\times 10^{19}$~eV and for proton primaries,
the dipole position should be aligned to the one observed in the CMB
within about 10$^\circ$. As a technical point that does not affect
qualitatively our estimate we note that the GMF is fixed with
respect to the galactic frame, not to the solar system one. Thus,
the CCG dipole is deflected by the GMF ``boosted'' by our relative
motion in the Galaxy. Similar considerations would apply to the
effect of possible diffuse fields in the local group of galaxies,
which is moving with respect to the CMB with $u_{\rm LG}\simeq
630\,$km$\,$s$^{-1}$~\cite{Eidelman:2004wy}.

{\it (iii)} Observing the CCG feature at only one energy provides
combined information on the intervening GMF and the charge of the
cosmic ray primaries. However, observations at two or more energies
break this degeneracy. For example, the determination of the primary
charge is straightforward as long as the cosmic rays propagate in
the quasi-ballistic regime and a single primary species dominates
the flux.  Denoting by $\hat{\delta}^{\rm CMB}$ the location of the
dipole in the CMB, and by $\hat{\delta}_1^{\rm CCG}$ and
$\hat{\delta}_2^{\rm CCG}$ the  dipole location measured via the CCG
effect at two energies with $E_1<E_2$, the primary charge $Q$ is
\begin{equation}
Q = \frac{\theta(\hat{\delta}^{\rm CMB},\hat{\delta}_1^{\rm
CCG})}{\theta(\hat{\delta}^{\rm CMB},\hat{\delta}_2^{\rm CCG})} \,,
\end{equation}
where $\theta$ is the angular distance.

{\it (iv)} Moving to lower energies, the anisotropy due to the CCG
effect should disappear as soon as galactic UHECRs start to
dominate. Relatively large anisotropies connected to an increased
source density in the disc or towards the galactic center are
expected to turn on somewhere between $10^{17}$~eV and the
ankle~\cite{Candia03}. The disappearance of the CCG anisotropy and
its replacement by galactic anisotropies is therefore an indicator
for the transition between galactic and extragalactic cosmic rays.

{\it (v)} Moving to sufficiently high energies,  $\lambda_{\rm hor}$
decreases and anisotropies due to local inhomogeneities in the
distribution of sources are expected to dominate. For protons, local
anisotropies should become important around 4--5$\times
10^{19}$~eV~\cite{Cuoco:2005yd}. Thus the decrease of the CCG
anisotropy with increasing energy is connected both to the amount of
inhomogeneity in the source distribution of UHECRs and to the
energy-loss horizon of the UHECR primary. This is also the energy
range where local motions, like the one towards the ``Great
Attractor'', might play an important role (for a review on motions
on large scales, see~\cite{Burstein90}).

\section{Detectability}
Is it possible for present experiments to detect a 0.6\% dipolar
anisotropy in the UHECR flux? In a sample of $N$ events, typical
fluctuations are of the order of $\sqrt{N}$. Thus a 0.6\% level
sensitivity is only reached for $\sqrt{N}/N\simeq 0.006$ or $N\simeq
3\times 10^4$ events. Reference~\cite{Mollerach:2005dv} gave an
empirical fit for the expected error $\sigma_A$ in the determination
of the amplitude of a dipole anisotropy as function of the event
number $N$ and the declination $\delta$ of the dipole vector,
\be \sigma_A = \sqrt{\frac{3}{N}}\left(1+0.6 \sin^3\delta\right),
\label{simulfit} \ee where a detector located at the PAO site and a
maximum zenith angle of 60$^\circ$ were assumed. Equation
(\ref{simulfit}) implies that a 3$\,\sigma$ detection of a 0.6\%
anisotropy requires of the order of $10^6$ events. Working from
January 2004 to June 2005, the PAO has reached a cumulative exposure
of 1750 km$^2$ sr yr, observing 1216 events in a 0.1 dex energy bin
around 10$^{18.55}$~eV~\cite{POAspec}. Once completed, the total
area covered will be around 3000~km$^2$, so that around 2000 events
per year should be collected close to the ankle. At this energy, one
can expect only a 1$\sigma$ hint in a decade of working time.
However, systematic errors are still quite large, and a shift in the
energy scale could significantly modify the flux estimate. More
importantly, since the UHECR spectrum is steep, better detection
possibilities are at lower energies, say between 10$^{17}$ and
10$^{18}$~eV, which will be explored in the near future by the PAO
and especially the TA thanks to the lower energy threshold. For
comparison, the 4\% anisotropy pattern found by the AGASA
collaboration around 10$^{18}\,$eV was based on about 284,000 events
collected after standard cuts at energies above
10$^{17}\,$eV~\cite{Teshima:2001me}. While a clear detection of the
CCG effect above the ankle will probably require future UHECR
observatories (see e.g.~\cite{JEM-EUSO}), the PAO and TA have the
realistic chance to disprove scenarios where the transition to
extragalactic cosmic rays happens below 10$^{18}\,$eV.
\section{The CCG effect and the diffuse $\gamma$-ray background}%
The CCG effect should be present in any cosmological diffuse
background. Thus one might wonder if the previous considerations
apply to other diffuse fluxes of interest in high energy
astroparticle physics. Such a case is the diffuse $\gamma$-ray
background, that may offer interesting detection perspectives of the
CCG effect, too. This background is a superposition of all
unresolved sources emitting $\gamma$-rays in the Universe and
provides thus an interesting signature of energetic phenomena over
cosmological time-scales. While a clear detection of this background
has been reported by the EGRET mission~\cite{Sreekumar:1997un}, its
origin is still uncertain. The original analysis of the EGRET data
derived an intensity spectrum of the unresolved flux in the GeV
region $I_\gamma\simeq 1.4\times 10^{-6}(E/{\rm GeV})^{-2.1}{\rm
cm}^{-2}{\rm s}^{-1}{\rm sr}^{-1}{\rm
GeV}^{-1}$~\cite{Sreekumar:1997un}. From this flux and the specifics
of the GLAST satellite experiment (a roughly energy independent
effective area $A_{\rm eff}=10^4$~cm$^2$ and a field of view of
$\Omega_{\rm fov}=2.4$~sr \cite{GLAST}) we can estimate the number
of events $N_\gamma$ above the energy $E_\gamma$ to be collected in
the time $t$ as
\begin{equation}
N_\gamma = A_{\rm eff}\,\Omega_{\rm fov}\, t
\int_{E_\gamma}^\infty{\rm d}E\,I_\gamma\simeq 9\times 10^5
\left(\frac{t}{{\rm yr}} \right) \left(\frac{E_\gamma}{{\rm
GeV}}\right)^{-1.1} .
\end{equation}
Such a relatively large statistics would allow one to detect the
signal at GeV energies at $3\,\sigma$ confidence level in about 1
year. The predicted amplitude of $(2+2.1)\,u\simeq 0.5\%$ (see
Eq.~(\ref{Ianisamplitude})) and the alignment of the dipole with the
CMB provide a smoking gun for the detection of the CCG effect. Note
that this signature would be useful to assess in a robust way the
cosmological (as opposed to the galactic) fraction of the diffuse
$\gamma$-ray background. Since the $\gamma$-ray flux after
extracting pointlike sources still contains a sizeable galactic
contamination, it is at present necessary to model the galactic
foreground. This foreground subtraction is however a delicate issue
as shown e.g.\ in the recent reanalysis of the EGRET data
in~\cite{Strong:2004ry}: Using a revised model for the galactic
propagation of cosmic rays, the deduced extragalactic spectrum was
estimated to be lower and with a different spectral shape than the
one reported in~\cite{Sreekumar:1997un}. Obviously, the CCG effect
provides a powerful, complementary tool in these analyses that are
for instance crucial in the detection of the putative diffuse
$\gamma$-ray signal from dark matter annihilations.

\section{Conclusions}%
We have argued that the cosmological Compton-Getting effect is a
powerful diagnostic tool for the study of UHECR physics, in
particular as a probe of the transition from galactic to
extragalactic cosmic rays and of the charge of the UHECR primaries.
Although challenging, the detection of the CCG effect appears within
reach with present detectors at least at energies below the ankle.
If UHECRs are of cosmological origin (i.e., they come from within an
energy-loss horizon of Gpc scale), the CCG effect should be the most
prominent large-scale anisotropy, similar to the case of the CMB.
The detection of a significantly larger dipole or of higher moments
with comparable size at energies around or above $10^{19}$~eV would
be difficult to explain, unless peculiar local sources are important
for UHECRs. The information encoded in the CCG effect motivates
serious experimental efforts towards its detection. A similar effect
is expected in the diffuse $\gamma$-ray background, with excellent
perspectives for its detection by GLAST. This signature might be
very useful in the difficult task to assess the cosmological
fraction of the measured  $\gamma$-ray background, which in turn
might indirectly affect the constraints on UHECRs physics as well.

\section*{Acknowledgments}%
We are grateful to G.~Raffelt for reading the manuscript. PS thanks
NTNU for hospitality during the initial phase of this work and
acknowledges the support by the Deut\-sche
For\-schungs\-ge\-mein\-schaft under grant SFB 375 and by the
European Network of Theoretical Astroparticle Physics ILIAS/N6 under
contract number RII3-CT-2004-506222.



\begin{thebibliography}{00}

\bibitem{reviews}
For recent reviews see e.g.\ M.~Nagano and A.~A.~Watson,
``Observations and implications of the ultrahigh-energy cosmic
rays,'' Rev.\ Mod.\ Phys.\  {\bf 72}, 689 (2000);
J.~W.~Cronin, ``The highest-energy cosmic rays,'' astro-ph/0402487;
M.~Kachelrie\ss, ``Status of particle physics solutions to the UHECR
puzzle,'' Comptes Rendus Physique {\bf 5}, 441 (2004)
[hep-ph/0406174].

\bibitem{PAO}
J.~W.~Cronin, ``Summary Of The Workshop,'' Nucl.\ Phys.\ Proc.\
Suppl.\  {\bf 28B}, 213 (1992).

\bibitem{TA}
M.~Fukushima, ``Telescope array project for extremely high energy
cosmic rays,'' Prog.\ Theor.\ Phys.\ Suppl.\  {\bf 151}, 206
(2003).

\bibitem{Hillas}
A.~M.~Hillas, ``The Origin Of Ultrahigh-Energy Cosmic Rays,'' Ann.\
Rev.\ Astron.\ Astrophys.\  {\bf 22}, 425 (1984).

\bibitem{ankle}
C.~T.~Hill, D.~N.~Schramm and T.~P.~Walker, ``Implications Of The
Ultrahigh-Energy Cosmic-Ray Spectrum Observed By  The Fly's Eye
Detector,'' Phys.\ Rev.\ D {\bf 34}, 1622 (1986);
J.~P.~Rachen, T.~Stanev and P.~L.~Biermann, ``Extragalactic
ultrahigh-energy cosmic rays. 2. Comparison with experimental
data,'' Astron.\ Astrophys.\  {\bf 273}, 377 (1993)
[astro-ph/9302005];
J.~N.~Bahcall and E.~Waxman, ``Has the GZK cutoff been
discovered?,'' Phys.\ Lett.\ B {\bf 556}, 1 (2003) [hep-ph/0206217];
T.~Wibig and A.~W.~Wolfendale, ``At what particle energy do
extragalactic cosmic rays start to predominate?,'' J.\ Phys.\ G {\bf
31}, 255 (2005) [astro-ph/0410624].


\bibitem{dip}
V.~Berezinsky, A.~Z.~Gazizov and S.~I.~Grigorieva, ``On
astrophysical solution to ultra high energy cosmic rays,''
hep-ph/0204357;
``Propagation and Signatures of Ultra High Energy Cosmic Rays,''
Nucl.\ Phys.\ Proc.\ Suppl.\  {\bf 136}, 147 (2004)
[astro-ph/0410650];
``Dip in UHECR spectrum as signature of proton interaction with
CMB,'' Phys.\ Lett.\ B {\bf 612} (2005) 147 [astro-ph/0502550].

\bibitem{Berezinsky:2004wx}
V.~S.~Berezinsky, S.~I.~Grigorieva and B.~I.~Hnatyk, ``Extragalactic
UHE proton spectrum and prediction for iron-nuclei flux  at
10**8-GeV to 10**9-GeV,'' Astropart.\ Phys.\  {\bf 21}, 617 (2004)
[astro-ph/0403477].


\bibitem{ZeV}
P.~Blasi, R.~I.~Epstein and A.~V.~Olinto, ``Ultra-high energy cosmic
rays from young neutron star winds,'' Astrophys.\ J.\  {\bf 533},
L123 (2000) [astro-ph/9912240].

\bibitem{chemie}
A.~A.~Watson, ``The mass composition of cosmic rays above
$10^{17}$~eV,'' Nucl. Phys. Proc. Suppl. {\bf 136}, 290 (2004)
[astro-ph/0408110].


\bibitem{Candia03}
J.~Candia, S.~Mollerach and E.~Roulet, ``Cosmic ray spectrum and
anisotropies from the knee to the second knee,'' JCAP {\bf 0305},
003 (2003) [astro-ph/0302082] and references therein.

\bibitem{Ber90}
V.~S.~Berezinsky {\it et al.},
``Astrophysics of Cosmic Rays,'' (North-Holland, Amsterdam, 1990).


\bibitem{CGpaper}
A.~H.~Compton and I.~A.~Getting, ``An Apparent Effect of Galactic
Rotation on the Intensity of Cosmic Rays,'' Phys.\ Rev. {\bf 47},
817 (1935).

\bibitem{Maoz:1993uq}
E.~Maoz, ``The Expected dipole in the distribution of cosmological
gamma-ray bursts,'' Astrophys.\ J.\  {\bf 428}, 454 (1994)
[astro-ph/9307039].

\bibitem{Eidelman:2004wy}
  S.~Eidelman {\it et al.}  [Particle Data Group],
  Phys.\ Lett.\ B {\bf 592}, 1 (2004).

\bibitem{ns}
P.~Blasi and D.~De Marco, ``The small scale anisotropies, the
spectrum and the sources of ultra  high energy cosmic rays,''
Astropart.\ Phys.\  {\bf 20}, 559 (2004) [astro-ph/0307067];
M.~Kachelrie{\ss} and D.~Semikoz, ``Ultra-high energy cosmic rays
from a finite number of point sources,'' Astropart.\ Phys.\  {\bf
23}, 486 (2005) [astro-ph/0405258].

\bibitem{Sigl:2004yk}
G.~Sigl, F.~Miniati and T.~A.~En\ss lin, ``Ultra-high energy cosmic
ray probes of large scale structure and magnetic fields,'' Phys.\
Rev.\ D {\bf 70}, 043007 (2004) [astro-ph/0401084].

\bibitem{Dolag:2004kp}
K.~Dolag, D.~Grasso, V.~Springel and I.~Tkachev, ``Constrained
simulations of the magnetic field in the local universe and the
propagation of UHECRs,'' JCAP {\bf 0501}, 009 (2005)
[astro-ph/0410419].

\bibitem{Kachelriess:2005qm}
M.~Kachelrie{\ss}, P.~D.~Serpico and M.~Teshima, ``The galactic
magnetic field as spectrograph for ultra-high energy cosmic rays,''
Astropart.\ Phys., in press [astro-ph/0510444].

\bibitem{Cuoco:2005yd}
A.~Cuoco  {\it et al.}, ``The footprint of large scale cosmic
structure on the ultra-high energy cosmic ray distribution,'' JCAP
{\bf 01}, 009 (2006) [astro-ph/0510765];
M.~Kachelrie{\ss} and D.~V.~Semikoz, ``Clustering of ultra-high
energy cosmic ray arrival directions on medium scales,'' Astropart.\
Phys., {\bf 26} (1), 10 (2006) 
[astro-ph/0512498].


\bibitem{Burstein90}
D.~Burstein, ``Large-scale motions in the Universe: a review,'' Rep.
Prog. Phys. 53 421 (1990).

\bibitem{Mollerach:2005dv}
S.~Mollerach and E.~Roulet, ``A new method to search for a cosmic
ray dipole anisotropy,'' JCAP {\bf 0508}, 004 (2005)
[astro-ph/0504630].

\bibitem{POAspec}
See \texttt{www.auger.org/icrc2005/spectrum.html}.

\bibitem{Teshima:2001me}
M.~Teshima {\it et al.}, ``Anisotropy of cosmic-ray arrival
direction at 10**18-eV observed by AGASA,''
in Proc.27th International Cosmic Ray Conference (ICRC 2001),
Hamburg, Germany, 7-15 Aug 2001.

\bibitem{JEM-EUSO}
Talk by F. Kajino at the 4$^{\rm th}$ Korean Astrophysics Workshop,
``JEM/EUSO project to study extreme universe by large and wide-angle
telescope'' available on-line at
\texttt{http://sirius.cnu.ac.kr/kaw4/}.

\bibitem{Sreekumar:1997un}
P.~Sreekumar {\it et al.}  [EGRET Collaboration], ``EGRET
observations of the extragalactic gamma ray emission,'' Astrophys.\
J.\  {\bf 494}, 523 (1998) [astro-ph/9709257].

\bibitem{GLAST}
See \texttt{http://www-glast.slac.stanford.edu/}.

\bibitem{Strong:2004ry}
A.~W.~Strong, I.~V.~Moskalenko and O.~Reimer, ``A new determination
of the extragalactic diffuse gamma-ray background from EGRET data,''
Astrophys.\ J.\  {\bf 613}, 956 (2004) [astro-ph/0405441].

\end{thebibliography}
\end{document}